\documentclass[aip,jcp,amsmath,amssymb,longbibliography,reprint]{revtex4-2}
\usepackage[utf8]{inputenc}
\usepackage[T1]{fontenc}
\usepackage{graphicx}
\usepackage{booktabs}
\usepackage[normalem]{ulem}
\usepackage{nicefrac}
\usepackage{microtype}
\usepackage{float}
\usepackage{bm}
\usepackage{color}
\usepackage{lipsum}
\setlipsum{
  par-before = \begingroup\color{blue},
  par-after  = \endgroup
}
\usepackage{tikz}
\usetikzlibrary{external}
\usepackage{hyperref} 
\begin{document}
\title{Toward high-throughput, low-latency structural biology using light, Part I: Protein-ligand pose determination}
\author{Brian K. Petkov}
\email{brian@junction.bio}
\affiliation{Junction Bioscience, New York, New York, USA}
\author{David R. Klug}
\affiliation{Junction Bioscience, New York, New York, USA}
\affiliation{Imperial College London, London, United Kingdom}

\begin{abstract}

We demonstrate for the first time that an optical spectroscopy---electron-vibration-vibration two-dimensional infrared spectroscopy (EVV 2DIR)---has sufficient structural competence to enable the reconstruction of the three-dimensional geometries of drug-protein binding at the level needed for practical drug discovery applications. Klug \textit{et al.} previously demonstrated experimental detection of drug binding to a kinase using EVV 2DIR. Here we use a quantum mechanical approach to calculate and assign elements of EVV 2DIR spectra for the spectral region 1200-1800 cm$^{-1}$ / 1100-1800 cm$^{-1}$ of 92 protein-ligand complexes. This data set includes ligand-target poses drawn from existing crystal structures and ligand-target poses created by docking and cofolding models. We find that binding-dependent spectral features are robust, interpretable, and consistent across poses in a given binding mode, and distinct for each unique hydrogen bonding geometry. We construct a spectral distance metric to distinguish poses that conform to the ligand-target hydrogen-bonding pattern of the crystal structure from those that do not. These results show that, given a set of predicted ligand-target poses containing the correct geometry, the hydrogen-bonding pattern of the valid pose may be rapidly identified by comparing simulated and experimental EVV 2DIR spectra. Finally, we discuss the practical implications for a new, high throughput structural biology for drug discovery that will be the subject of forthcoming manuscripts, noting that the base technologies to perform EVV 2DIR quickly and efficiently already exist and will enable a novel structure to be determined in only minutes for each sample. Moreover this can be achieved without the need for crystallization, isotopic labeling, or low temperatures, non-destructively using picogram quantities of material.

\end{abstract}


\maketitle

\section{Introduction}

The rate and efficiency with which new structures for drug discovery can be solved is severely restricted by the throughput and latency limitations of the dominant analysis techniques of x-ray crystallography and cryo-electron microscopy. These techniques determine atomic positions from electron densities of crystallized or cryo-frozen samples, respectively. A third less widely used approach involves measurement of nuclear spin couplings through a suite of interrelated nuclear magnetic resonance spectroscopy experiments. These generally require both isotopic labeling and relatively large quantities of material to obtain a structural model. The Protein Data Bank doubles in size roughly every seven years, reflecting both a significant scientific infrastructure achievement and the ongoing challenge of real-world structure determination\cite{Berman2000,Burley2025,RCSBGrowth}. Conversely, optical spectroscopies are commonly used for the detection and quantification of protein-ligand binding but have lacked the structural competence to determine the pose of a ligand-protein complex. Two-dimensional infrared spectroscopy is emerging \cite{Woutersen2026} as an approach for capturing elements of biomolecular structural information albeit typically without explicit geometric resolution. In this publication we show that the EVV variant of 2DIR spectroscopy has particular potential as a structural biology tool and we demonstrate that it contains information which is more than sufficient to competently identify a correct ligand-protein pose with a high degree of confidence. \par 
\begin{figure*}[t]
    \centering
    \includegraphics[width=1.0\linewidth]{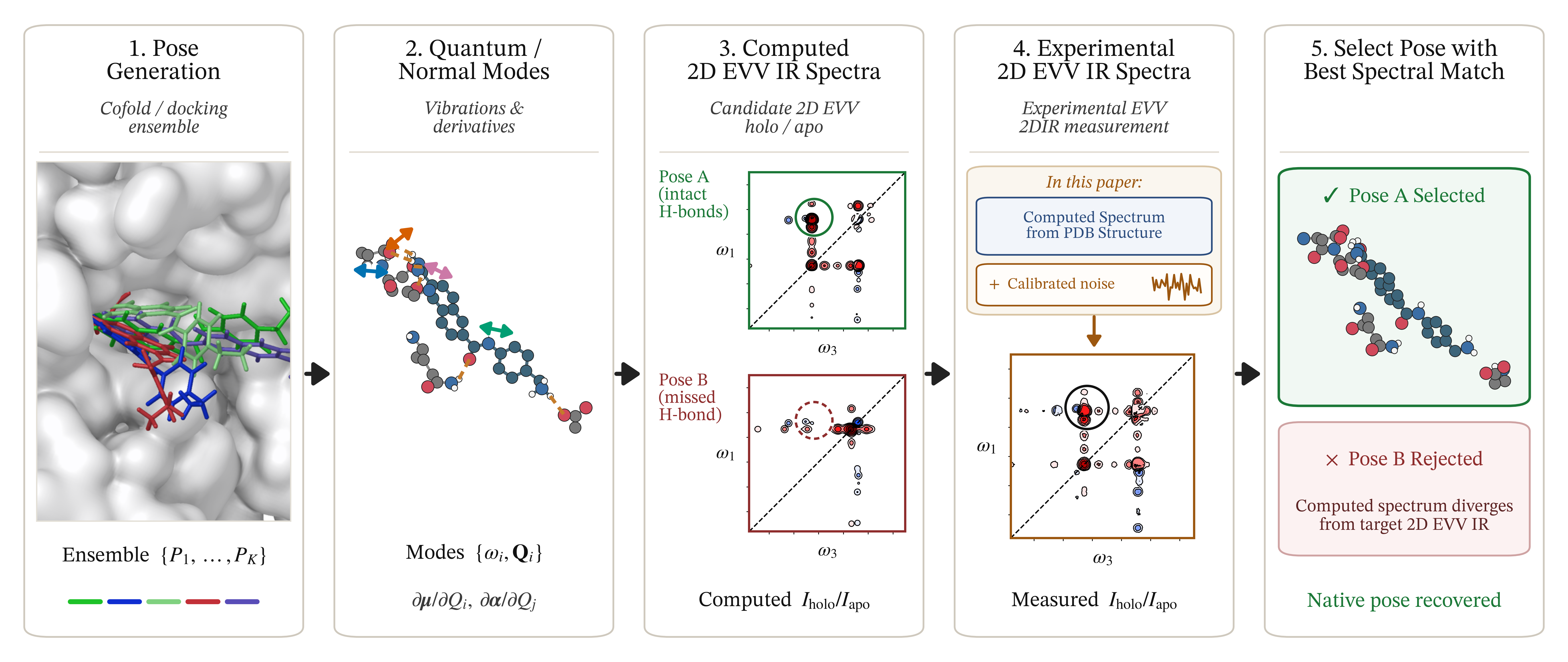}
    \caption{\textbf{Schematic of the proposed EVV 2DIR pose determination strategy.} (1) An ensemble of poses of the ligand in the binding are generated using physics-based or cofolding approaches. (2) Normal modes and dipole/polarization derivatives are computed from first-principles quantum physics or neural network potentials. (3) EVV 2DIR spectra are computed for the poses in the ensemble. (4) EVV 2DIR spectra are measured as a ground truth reference for the sample. In this proof of concept, we instead utilize computed spectra from existing Protein Data Bank structures as ground truth. (5) Predicted pose spectra are compared to the ground truth spectrum to select the best match or rule out all hypotheses in the ensemble.}
    \label{fig:placeholder}
\end{figure*}

Electron-vibration-vibration 2DIR is a variant first experimentally described by Zhao and Wright\cite{Zhao1999,Zhao2000,Zhao2000ApplSpectrosc} as doubly-vibrationally-enhanced four-wave mixing and later extended by Klug to determine the geometry of molecular complexes in a simple system\cite{Guo2009,Guo2012}, to detect the binding of a ligand to a target protein\cite{Sowley2019}, and to identify folding behavior in quadruplex DNA\cite{Soenarjo2023}. Conceptually, the technique uses a combination of short infrared and visible light pulses to measure quantum couplings between molecular vibrational modes. The EVV variant of 2DIR spectroscopy has particular sensitivity to both through-space electrical and through-bond mechanical coupling and typically identifies $\sim$100 measurable couplings for large biomolecules within the spectral range used in this paper\cite{Soenarjo2023,Sowley2019}. Early work by Guo \textit{et al.} established the forward-mapping between molecular geometry and EVV 2DIR spectra for a mixture of two aromatic small molecules and demonstrated that experimental spectra contain geometry-dependent peaks in agreement with spectra computed from both density functional theory\cite{Guo2011,Guo2012} and more advanced quantum calculations\cite{Guo2011,Guo2012}. Later work extended EVV 2DIR to detection in biochemical systems, demonstrating an accuracy of quantification of $\sim$100\% and a lower limit of detection of $\sim$1\% for tyrosine nitration in a peptide\cite{RezendeValim2014}, and tracking the salt-dependent folding of G4-quadruplex DNA\cite{Soenarjo2023}, including the ability to detect a small (4\%) sub-population within the sample. An important feature of these data are a dynamic range of $\sim$500-1000\% and noise below 10\% of each feature size, even toward the bottom of the dynamic range. This means that both weak and strong spectral features can be recovered with good fidelity. Equally, the ability to reliably measure small changes caused by a controlled perturbation enables features of interest to be revealed and distinguished from features unchanged by the perturbation.
The perturbation approach was used to study drug-protein interactions\cite{Sowley2019}. In this case the addition of a ligand to a target protein perturbs the intramolecular vibrational couplings and forms new intermolecular vibrational couplings. The effect of these perturbations become apparent by measuring the difference or ratio of the Holo/Apo spectrum. In the first published application to small molecule drug discovery, Sowley \textit{et al.} demonstrated that EVV 2DIR spectra contain sufficiently numerous and well-resolved binding-dependent peaks which can be distinguished from trivial additional peaks due to intramolecular couplings within the ligand\cite{Sowley2019}. Titration confirmed the assignment of binding-dependent peaks to support the detection of the binding of SU5402 to FGFR1 kinase and distinguish binding-dependent peaks from intramolecular couplings of the ligand. While steps toward assignment of peaks from geometry were taken, the paper did not explore a full inverse-mapping from spectrum to structure. 
The quality of EVV 2DIR data has for some time been sufficient in principle to enable a strategy of structural hypothesis acceptance/rejection, as demonstrated in the G4 DNA publication\cite{Soenarjo2023}. However, to make this a practicable technique in, for example, ligand pose determination for the general protein-ligand case, there is a need to be able to generate large numbers of structural hypotheses and to accurately calculate the EVV 2DIR spectra of these, so that they can be compared against a target spectrum. The target spectrum is envisaged as an experimentally-determined spectrum of the Holo/Apo ligand-protein system with dynamic range and noise characteristics as described above. Advances in neural network potentials have enabled rapid, efficient and accurate generation of EVV 2DIR spectra from a structural hypothesis, enabling acceptance or rejection of the hypothesis based on appropriate criteria. It is this structure determination strategy which we present here.
 
\begin{figure*}
    \centering
    \includegraphics[width=1.0\linewidth]{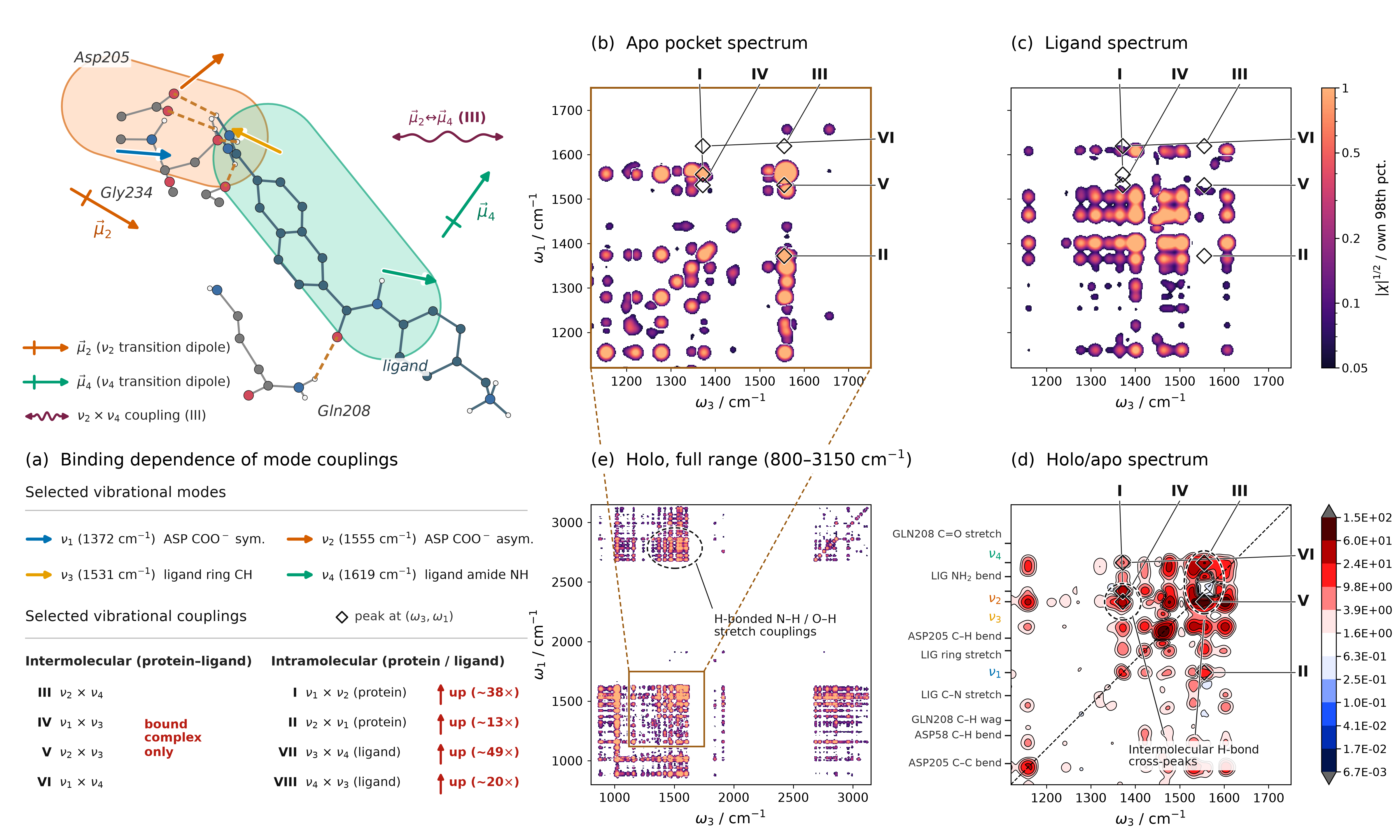}
    \caption{\textbf{Illustrative EVV 2DIR spectral simulation of small molecule ligand and protein complex (PDB 9GRR).} \textbf{(a)} Selected couplings between the transition dipoles $\mu$ of vibrational modes $\nu$ of the ligand-protein complex. Binding induces changes in coupled intramolecular ligand and protein modes and introduces new intermolecular couplings. A subset of modes are depicted here for illustrative purposes. In the results presented here we do not make a dipole approximation, but the electric fields are still directional and generate geometric sensitivity. \textbf{(b,c)} Apo protein and ligand spectra presenting features originating from intramolecular vibrational mode coupling. \textbf{(d)} Holo/apo ratio spectrum obtained by division of the bound-state spectrum by the apo spectrum. Drug binding modulates the intensity of associated coupling peaks over a $\sim$1000 dynamic range. Elements in this spectrum indicate a change in intermolecular or intramolecular vibrational mode coupling upon binding. Spectral features can be traced back to the originating vibrational couplings and arise directly from binding site geometry. \textbf{(e)} Expanded-range EVV 2DIR spectrum of bound ligand-protein complex, displaying in context the spectral region addressed in this work.}
    \label{fig:placeholder}
\end{figure*}
 \par
In this paper we identify the correct pose---defined here as the pose with hydrogen bonding geometry closest to that of the corresponding Protein Data Bank crystal structure---from a set of physics-based docking and cofolding-generated candidate poses using only the computed EVV 2DIR spectra as a discriminator. The calculated spectrum of the crystal structure pose is used as the target spectrum for demonstration purposes, but in a fully functioning high throughput industrial setting, the target spectrum would be provided by experimental measurement of the EVV 2DIR spectrum of the protein-ligand system. In order to demonstrate the practicality of this approach, we furthermore add realistic levels of noise to the simulated spectra in order to take account of the finite S/N of a real experimental system. The spectral frequency window, dynamic range, and peak width are confirmed to conform to that achieved by experimental EVV 2DIR for biochemical systems\cite{Sowley2019, Soenarjo2023}. We also touch upon several unique additional advantages of this approach. These include the ability to detect and characterize hydrogen bonds themselves rather than infer their existence~\cite{Sowley2019}, the ability to refine poses to sub-angstrom resolution, the potential for structure determination of membrane-bound and non-crystallizable protein-ligand complexes, the avoidance of complex sample preparation such as labeling, crystallization or low temperatures and the ability to generate data of sufficient S/N and dynamic range in only a few minutes for each sample. Finally, as demonstrated in a previous publication\cite{Guo2012}, the exquisite sensitivity of EVV 2DIR features to geometry implies the ability to discriminate between structures at the sub-angstrom level. Our calculations in this paper of EVV 2DIR spectra for drug-ligand poses and their sensitivity to sub-angstrom structural changes supports this conclusion.
\section{Materials And Methods}
\begin{figure*}
    \centering
    \includegraphics[width=1\linewidth]{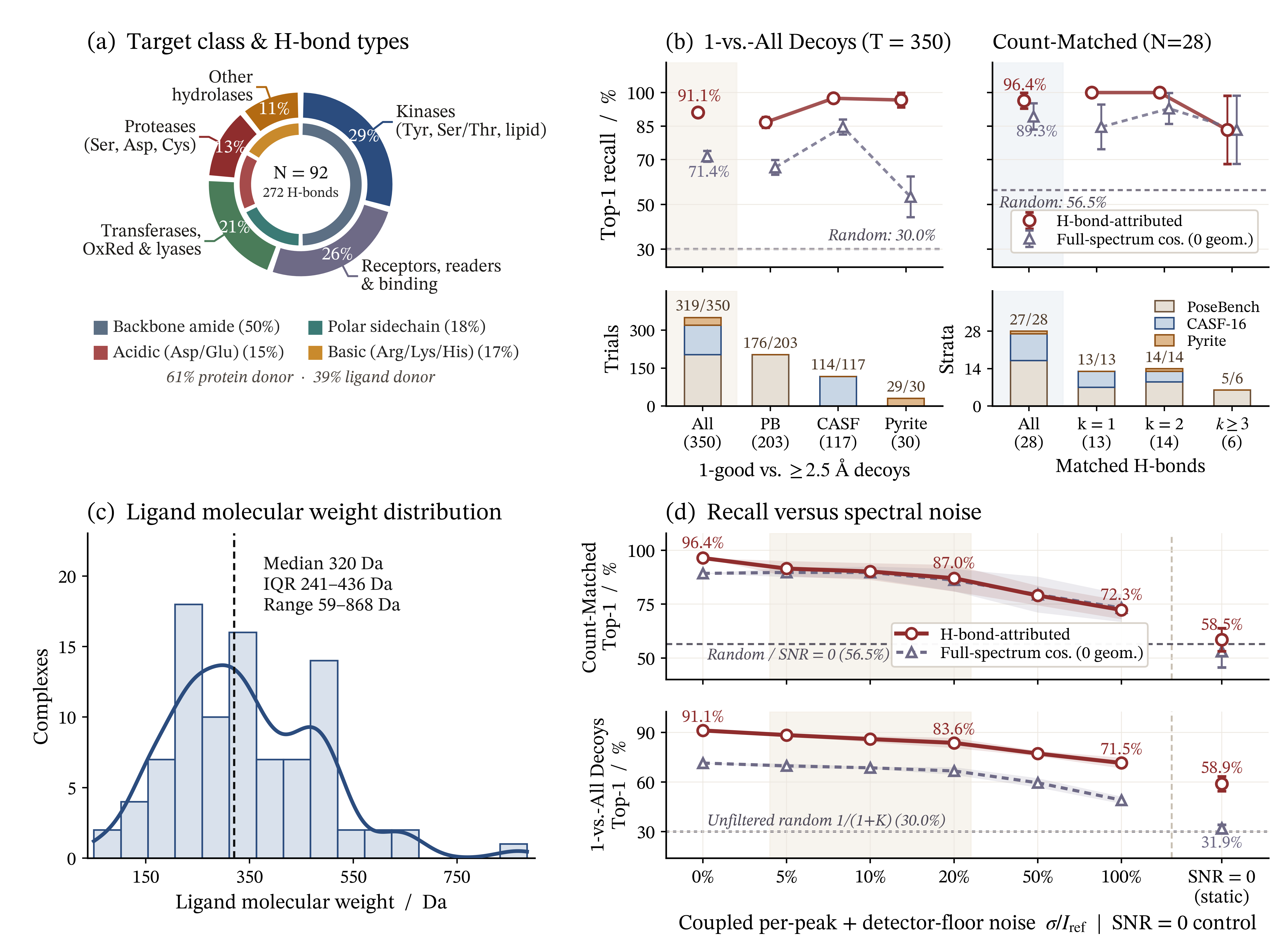}
    \caption{\textbf{Recall of correct ligand pose using EVV 2DIR as a discriminator.} \textbf{(a)} Target and binding-site statistics of ligand-protein complexes used in pose-selection study \textbf{(b)} Top-1 Recall of ligand pose by originating dataset. PoseBench and Pyrite correct/incorrect poses are cofolded; CASF-16 poses are docked; PoseBench includes docked and cofolded poses. (Left) Top-1 pose recall by cohort (mean $\pm 1\,\mathrm{s.d.}$ across $n = 8$ independent Monte Carlo realizations). Random baseline is computed from the fraction of good poses to total poses for full and count-matched subsets. \textbf{(c)} Distribution of ligand molecular weight across the systems evaluated 
     \textbf{(d)} Spectral noise sensitivity and zero-signal ($\mathrm{SNR} = 0$) negative control. H-bond attributed recall utilizes a spectral metric that prioritizes spectral features attributed to intermolecular hydrogen bonds; full-spectrum cosine uses the full Holo/Apo spectrum. 100\% random noise corresponds to background of equal magnitude to a given peak (SNR = 1). A pure random noise spectrum was used as a zero-signal control.}
    \label{fig:placeholder}
\end{figure*}
 The protein–ligand pose dataset was assembled from public curated benchmarks: PoseBench 1.1~\cite{Morehead2026} (includes PoseBusters~\cite{Buttenschoen2024},
Astex Diverse~\cite{Hartshorn2007}, and DockGen-E~\cite{Corso2024} predictions from
AlphaFold 3~\cite{Abramson2024}, Boltz-1~\cite{Wohlwend2024}, Chai-1~\cite{ChaiDiscovery2024},
RoseTTAFold All-Atom~\cite{Krishna2024}, DiffDock-L~\cite{Corso2024}, and
AutoDock Vina~\cite{Trott2010,Eberhardt2021}), CASF-2016~\cite{Su2019}
rigid-receptor docking ensembles, and Pyrite~\cite{Srivastava2026} cofolded ensembles. These sets were filtered to exclude covalent complexes and elements beyond (H,C,N,O,F,P,S,Cl,Br,I). Systems with over 60 ligand heavy atoms or containing active-site metals, hemes, or secondary cofactors within 4.5 \AA\ of the primary ligand were also excluded. Systems were required to form at least one intermolecular hydrogen bond in the crystal reference. Reference complexes were protonated and hydrogens were relaxed at pH~7.4 using
Hydride~\cite{Kunzmann2022}, and each candidate pose was protonated identically to the reference using a topology-mapping approach. Poses with significant intramolecular or intermolecular clashes or bond length distortion prior to and following this protonation procedure were discarded. 

Intermolecular hydrogen bonds were provisionally assigned using ProLIF~\cite{Bouysset2021} donor--acceptor fingerprints controlling for resonance-equivalence. From the resulting clean candidate set, complexes were assembled into balanced multi-pose cohorts pairing diverse near-native poses that preserve the crystal hydrogen-bond network with RMSD-stratified non-native decoys with disrupted crystal hydrogen bonds. Each complex was truncated to the shared union of pocket residues participating in any crystal or candidate hydrogen bonds so that all holo and apo fragments within a system maintained identical atom composition and formal charge. 

EVV 2DIR spectra were calculated for the purely electrically coupled case\cite{Guo2009,Guo2011,Guo2012,Soenarjo2023}. In order to speed up the calculations, EVV 2D-IR spectra were simulated using a hybrid neural network potential (NNP) workflow that decouples harmonic normal-mode analysis from electrical response derivatives was used. Harmonic vibrational frequencies and normal-mode eigenvectors were computed at fixed
molecular charge and spin multiplicity using the FAIRChem UMA-M-1.1 interatomic
potential~\cite{Wood2025} (trained on the OMol25~\cite{Levine2025} datset at the
$\omega$B97M-V/def2-TZVPD level~\cite{Mardirossian2016,Rappoport2010}) via Cartesian
finite-difference Hessian evaluation. Modes within the 900--3200~cm$^{-1}$ frequency window were retained,
mass-weighted to orthonormality, and scaled by the $\omega$B97M-V fundamental factor of
0.958~\cite{Liang2023}. First- and second-order transition dipole derivatives
($\partial\mu/\partial Q_k$ and $\partial^2\mu/\partial Q_j\partial Q_k$) and
first-order polarizability derivatives ($\partial\alpha/\partial Q_k$) were then evaluated
at the same geometry using the equivariant MACE-MDP electrical model~\cite{Batatia2022,Gonnheimer2026}, which is trained specifically for prediction of $\mu$ and $\alpha$, but not the associated derivatives. Dipole and polarizability derivatives were obtained by projecting Cartesian
Jacobians and Hessians onto the mass-weighted UMA normal coordinates across all mode
pairs in the padded frequency window. Within the electrical-anharmonicity approximation used, orientationally averaged third-order susceptibility amplitudes for parallel ($\chi^{(3)}_{\mathrm{PPP}}$) and perpendicular ($\chi^{(3)}_{\mathrm{PPS}}$) polarization geometries were formed by contracting $\partial\bm{\mu}/\partial Q_a$, $\partial\bm{\alpha}/\partial Q_b$, and $\partial^2\bm{\mu}/\partial Q_a\partial Q_b$, coherently summed with separable 2D Gaussian lineshapes ($15~\mathrm{cm}^{-1}$ FWHM over $\omega_\alpha \in [1200, 1800]~\mathrm{cm}^{-1}$ and $\omega_\beta - \omega_\alpha \in [1100, 1800]~\mathrm{cm}^{-1}$), and squaring to obtain homodyne EVV intensities and regularized holo/apo ratio spectra. The magnitude of each Gaussian envelope was randomly scaled by 10\% on both holo and apo spectra, 10\% white Gaussian noise was applied to crystal holo and apo across the spectral grid, and a noise floor of 2$\%$ was applied to model experimental detection conditions and sensitivity. Eight random seeds were drawn independently, after which the full noised scoring pipeline was applied to all poses to compute recall. Samples spanning a range of 0\% to 100\% of peak magnitude were drawn to further model a range of experimental noise conditions, with 100\% noise corresponding to background of equal magnitude to a given spectral feature (SNR = 1).

To assign spectral features resulting specifically from a given intermolecular protein-ligand hydrogen bond, the predicted EVV 2DIR difference spectrum for each pose was decomposed by isolating the electrical coupling arising exclusively from the coupled motion of each donor-acceptor pair. This localizes each candidate hydrogen bond to a small set of up to 8 characteristic 50 x 50 cm$^{-1}$ regions defined by cross-peak frequencies in the pose Holo/Apo difference spectrum. The crystal spectrum was also assigned up to 10 regions based on peak intensity. Each characteristic region was then scored for shape and magnitude alignment with the corresponding crystal spectra so as to reward/penalize the presence/absence of similar spectral features between the two poses.

\section{Results And Discussion}

\begin{figure*}
    \centering
    \includegraphics[width=1.0\linewidth]{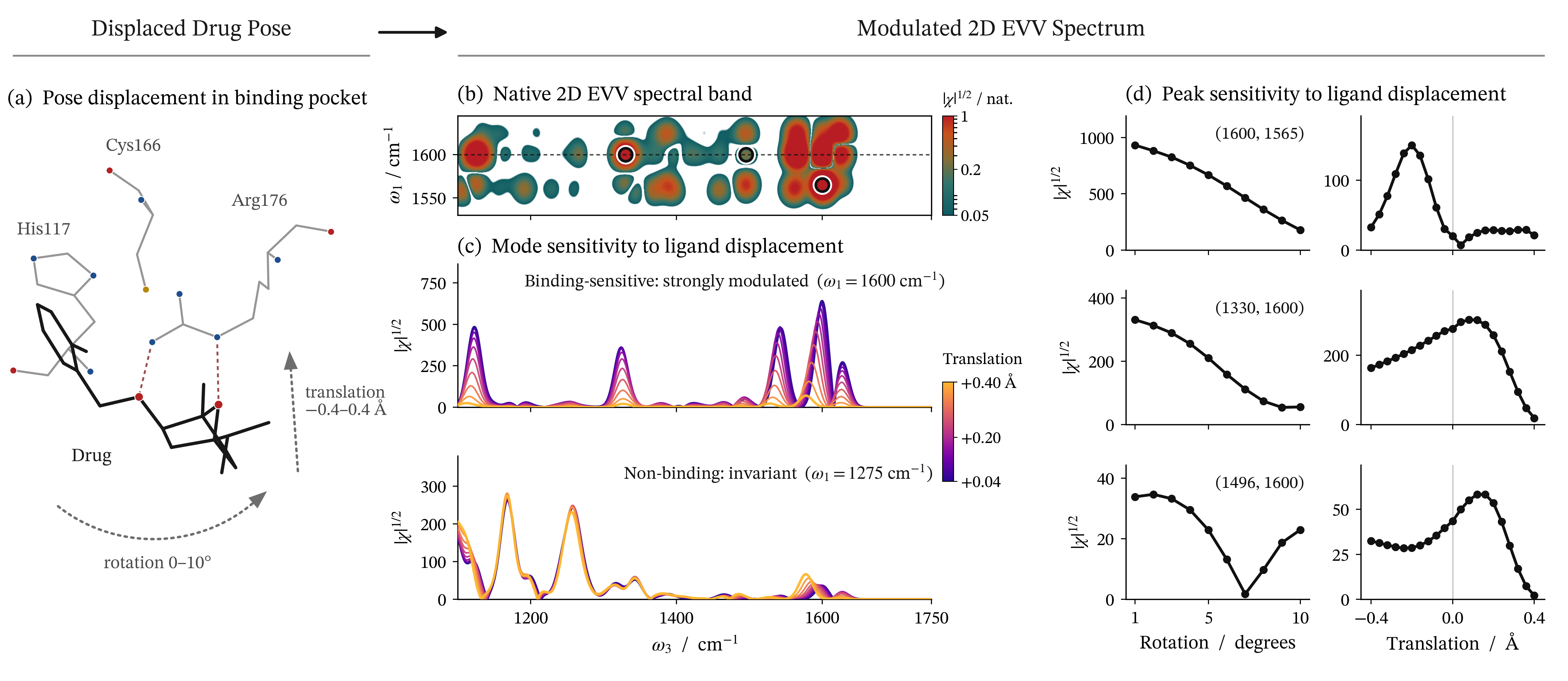}
    \caption{\textbf{Sensitivity of EVV 2DIR spectra to drug binding interface geometry.} \textbf{(a)} Ligand in binding pocket with axes of rigid rotation and translation depicted. Displaced intermediate poses not pictured. \textbf{(b)} Subregion of EVV 2DIR spectrum of protein-ligand complex, with three peaks circled at (1350,1600), (1510,1600), and (1600,1565) cm$^{-1}$. \textbf{(c)} (Top) 1D slice from 2D spectrum in (b) taken at $\omega_1 = 1600$ cm$^{-1}$. Ten slices are plotted with intensity dropping substantially with translation. (Bottom) 1D slice from 2D spectrum (range not pictured in (b)) taken at $\omega_1 = 1275$ cm$^{-1}$. Ten slices are plotted with intensity largely invariant to translation. \textbf{(d)} Intensity of peaks circled in (b) as a function of rigid rotation and translation coordinate.}
    \label{fig:placeholder}
\end{figure*}


The objective of this work is to establish a basis for using EVV 2DIR to determine the pose geometry for ligand-bound protein structures. In this publication we describe the particular properties of EVV 2DIR that enable it to be a structure-determining method and present the key results that support a robust mapping between molecular geometry and experimentally observable spectral features. Our focus here is on small molecule ligands. 

The enabling principles are as follows: i.) EVV 2DIR holo/apo spectral ratios contain numerous and sufficiently well-separated identifiable features that enable the quantification required for structure determination; ii.) Features originate from coupling of inter- and intramolecular vibrational modes and are a function of pose geometry; iii.) Mode couplings, and therefore EVV 2DIR features, are highly sensitive and undergo shifts of  frequency and intensity upon binding, including the creating of new couplings; iv.) Binding-associated spectral changes contain sufficient information to discriminate between poses; v.) EVV spectra in conjunction with a structural hypothesis-generation engine such as docking or cofolding allow for atomic determination of ligand-bound geometry by ruling in (out) structural hypotheses that do (do not) fit the target data in such a way that correctly identifies the hydrogen bonds between ligand and protein. 

The ability to detect ligand binding and assign at least one associated change in EVV 2DIR spectrum was reported previously\cite{Sowley2019}. The sensitivity of EVV 2DIR spectroscopy to molecular geometry was also demonstrated previously\cite{Guo2009,Guo2012, Soenarjo2023}. The missing elements toward a practicable high-throughput structure-determining tool were (A) the ability to calculate EVV 2DIR spectra with sufficient speed and accuracy to keep up with experimental data generation; and (B) A robust algorithm for interpreting hydrogen bond features of EVV 2DIR spectra so that the pose determination can be made on the basis of the correct number and identity of the hydrogen bonds present. Capability A was computationally intractable at scale until the recent advent of high performance and scalable GPU quantum chemistry codes and neural network potentials. Capability B required the generation of a large data set and the development and testing of a pose discrimination algorithm that is also resistant to experimental noise.

To evaluate whether EVV 2DIR can recall correct poses from a set without using any other structural priors, we benchmarked blind pose ranking across N = 92 diverse protein-ligand complexes. The structures used were curated by filtering publicly available benchmark sets for clashes and distorted geometries---common in raw docking and cofolding model outputs---that must be resolved prior to running normal mode and quantum calculations. In total, 785 candidate poses and 941 native-versus-decoy pairs were compared. A Top-1 recall of 91.1\% was achieved using the hydrogen-bond-attributed spectral distance metric described earlier. Recall is higher on the CASF-2016 (97.4\%) and Pyrite (96.7\%) than on the PoseBench 1.1 (86.7\%) set. Notably, the CASF-2016 poses are docked to a rigid receptor, whereas all of the Pyrite poses are independently cofolded using the Pearl model and allow for conformational differences in pocket residues versus the (blinded) crystal structure. Performance differences between PoseBench 1.1 and Pyrite---both cofolded---systems may be attributed to the fact that conformational differences in the protein pocket relative to crystal are much greater in the former. PoseBench 1.1 structures show sidechain shifts of 1.5 $\AA$ on average and as much as 7 $\AA$ among decoys, nearly 4x greater than the Pyrite subset. This represents a substantial induced fit challenge that we expect to be addressed by utilizing additional spectral features, expanding the spectral frequency range, and/or including the remaining non-hydrogen bonding pocket residues into the set used for spectral calculation. 

To further probe pose discrimination performance, we curated a count-matched subset of 28 systems requiring each pose (both native and decoy) to contain the same number of intermolecular hydrogen bonds, and found that recall remains high at 96.4\%. 11 systems contained pairs with an identical hydrogen bonding pattern to the crystal (same donor-acceptor pairing), representing a challenging case for pose discrimination. On this bond-matched subset, the EVV 2DIR approach was successful in recalling the pose with lowest RMSD to crystal in 10/11 cases using the hydrogen bond attributed spectral metric. This performance was higher than anticipated given that the hydrogen bond attributed spectral metric and protocol attempted here was not designed for \textit{refinement} of a pose geometry within a particular binding mode. Full-spectrum cosine similarity recalls all 11 bond-matched cases, suggesting that spectral features originating from other intermolecular interactions and contacts may be used to refine a pose once the correct hydrogen-bonding geometry has been established. This result is consistent with the higher recall ability of the hydrogen bond focused spectral metric observed when poses differ in their hydrogen bonding geometry, compared to full spectrum cosine similarity which assigns equal importance to features derived from non-binding intramolecular couplings. 

To explain this high recall we investigate the geometric sensitivity of the EVV 2DIR spectra in a sample (PDB 9GRR) protein-ligand system (Figure 4(c,d)). Rigid-body scans of ligand displacement along arbitrary clash-free rotational and translational axes demonstrate the varied binding-sensitivity of spectral features. Ligand translation in the binding pocket along $+0.04$ to $+0.40~\text{\AA}$ from the crystal pose produce pronounced amplitude modulation along the ($\omega_1 = 1600~\mathrm{cm}^{-1}$) binding-sensitive cross section, with amplitude falling orders of magnitude across the range at most frequencies. In contrast, spectral cross-sections through non-binding modes ($\omega_1 = 1275~\mathrm{cm}^{-1}$) remain virtually invariant across this range. Tracking individual features attributed to intermolecular coupling---such as $(1600, 1565)$, $(1330, 1600)$, and $(1496, 1600)~\mathrm{cm}^{-1}$---reveals steep, nonlinear sensitivity in response to single-degree rotations ($1\text{--}10^\circ$) and sub-angstrom translations. This example highlights the sensitivity of EVV 2DIR spectra to even sub-angstrom differences in binding mode, and is representative of the sensitivity observed across systems in this and prior work\cite{Guo2011, Guo2012}.


Together, these results establish that NNP-predicted EVV 2DIR spectra translate hydrogen-bond geometry into a noise-resilient spectroscopic fingerprint capable of resolving near-native protein--ligand poses across diverse drug-discovery targets. The spectral metric used here effectively recalls poses with similar intermolecular hydrogen bonding geometries to the crystal over poses bound with incorrect hydrogen bonding geometries. The metric is robust to modeled experimental noise, and spectra are calculated over an experimentally realistic frequency and dynamic range. Furthermore, the metric shows promise in refining poses within a given binding mode and hydrogen bonding configuration. The structure-solving strategy demonstrated here is enabled by the significant advances that have been made across a number of technologies in recent years\cite{Woutersen2026}. Firstly there is the EVV 2DIR spectroscopy itself, which is able to rapidly produce high information-content signal with high dynamic range, using small quantities of sample, at room temperature, without labeling, crystallization or other complex sample manipulation\cite{Donaldson2007,Donaldson2008,RezendeValim2014,Sowley2019}. Sensitivity and dynamic range are further improved by diode-pumped Yb:KGW amplifiers, which now operate reliably at high repetition rates and provide substantial gains in long-term stability and signal-to-noise\cite{Donaldson2023}. Secondly, there is the advance of quantum mechanical and neural network potential  methods\cite{Wood2025,Levine2025,Batatia2022,Gonnheimer2026} that enable EVV 2DIR spectra to be calculated both sufficiently accurately and rapidly enough to keep pace with the experimental throughput and provide the analysis of the experimental data. Finally, there is the advance in cofolding and docking methods that provide an ensemble of ligand-protein poses\cite{Abramson2024,Krishna2024,Wohlwend2024,ChaiDiscovery2024,Corso2024,Trott2010}, from which at least one may conform to the true pose attained when the ligand and protein are bound together. 


Structural validity and consistency limitations of docking, conformer generation, and cofolding models are important but addressable by sanitization prior to computing normal modes and derivatives. While not discussed here, the potential to extract experimental priors from EVV 2DIR spectra may be of utility in regimes where structural sampling is a limiting factor. Calculations above were limited to the immediate interacting residues in the binding pocket, which is a reasonable approximation given that measurement of holo/apo ratio spectra zeroes out non-binding protein response. Given that the addition of residues to the calculated binding pocket will increase the information density of the spectra, pose discrimination is likely to improve in turn given that both computed and experimental EVV 2DIR spectra present hundreds of well-separated peaks. Induced fit effects are partially addressed by the use of cofolding structures here that include conformational change in pocket residues, but we have omitted consideration of more extensive conformational change as a topic to be addressed in a future publication. The fidelity and speed of the quantum physics calculation is an important consideration for prospective deployment of the method. The neural network potentials applied in this work were selected for demonstration purposes given their high agreement with DFT, public availability, and relative speed~\cite{Wood2025,Levine2025,Batatia2022,Gonnheimer2026}.   

We have omitted solvent and salts from consideration altogether in this work. In practice, EVV 2DIR measurements are performed with the protein surrounded by water and aqueous salts at room-temperature conditions\cite{Sowley2019,Woutersen2026} resembling the native cellular environment. The measured spectral response will thus contain contributions from the solvation network, and the geometry of water-ligand and water-protein hydrogen bonding interactions may be approached in analogy to the strategy described herein for protein-ligand geometries. Results incorporating water interactions will be the subject of a future publication. This paper emphasizes intermolecular hydrogen bonds as a unique and concrete geometric feature mapped by EVV 2DIR spectra, but the technique is sensitive to mode couplings in general and prior works have mapped, for example, pi-stacking geometries to experimentally-resolved spectral features\cite{Guo2009}. Expansion to the full accessible spectral region, where detection of aliphatic modes has been experimentally demonstrated\cite{Donaldson2008} and may support determination of hydrophobic binding site geometries; and to additional spectral dimensions, will be presented in forthcoming work.

\section{Conclusion}
We have shown that EVV 2DIR spectra are a viable means of recalling the correct drug-bound complex geometry from a set of docking or cofolding predictions. Spectra are remarkably sensitive to geometry and present well-resolved, measurable features that can be used to resolve fine differences in geometry and hydrogen-bonding configuration. The strategy described in this publication represents the basis upon which EVV 2DIR spectroscopy is positioned to contribute to drug design and structural biology more broadly.

\section*{Acknowledgments} 
The cloud computing infrastructure used to perform docking, cofolding, and quantum spectral calculations was implemented by Janan Zhu. We acknowledge generous GPU compute support from Google Cloud Platform and thank Varsha Bhargava and Corey Bayne from the GCP team. We thank Matthew Gummow, Harrison Esterly, Rafal Wiewiora, Thomas Huber, Spencer Frazier, Thomas Sakmar, and Daniel Turner for helpful discussions. 

\section{Competing Interests}
The authors are affiliated with Junction Bioscience, a company with a commercial interest in electron-vibration-vibration two-dimensional infrared spectroscopy.

\bibliography{references}

\clearpage
\onecolumngrid
\appendix

\begin{center}
  {\LARGE\bfseries Supporting Information}\\[0.5em]
  {\large Toward high-throughput, low-latency structural biology using light, Part I: Protein-ligand pose determination}
\end{center}
\vspace{1.5em}

\setcounter{section}{0}
\setcounter{equation}{0}
\setcounter{figure}{0}
\setcounter{table}{0}
\setcounter{page}{1}

\renewcommand{\thesection}{S\arabic{section}}
\renewcommand{\thesubsection}{S\arabic{section}.\arabic{subsection}}
\renewcommand{\theequation}{S\arabic{equation}}
\renewcommand{\thefigure}{S\arabic{figure}}
\renewcommand{\thetable}{S\arabic{table}}
\renewcommand{\thepage}{S\arabic{page}}

\renewcommand{\theHsection}{S\arabic{section}}
\renewcommand{\theHequation}{S\arabic{equation}}
\renewcommand{\theHfigure}{S\arabic{figure}}
\renewcommand{\theHtable}{S\arabic{table}}

\section{Illustrative Comparison of EVV 2DIR Spectra for Pose Discrimination}
\label{sec:si-comp}

\begin{figure}[!htbp]
  \centering
  \includegraphics[width=\textwidth, height=0.65\textheight, keepaspectratio]{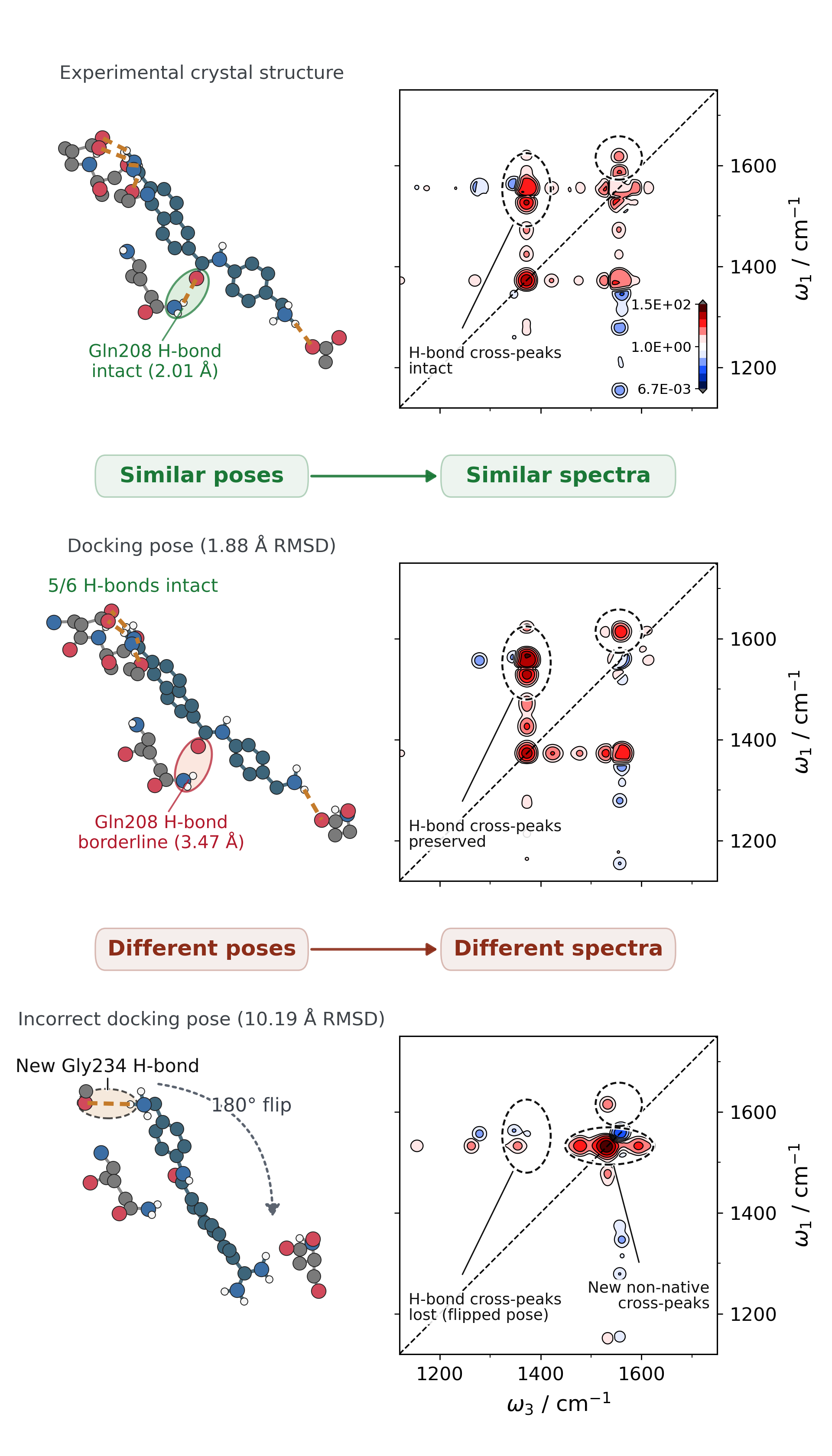}
  \caption{Comparison between EVV 2DIR spectra of PDB 1OWH pocket residues with 6 intermolecular hydrogen bonds between crystal and ligand (top), a docking pose that retains all but one borderline GLN208-ligand hydrogen bond (middle), and an incorrect docking pose that erroneously flips the ligand nearly 180 degrees relative to crystal (bottom). Spectra depict only a selection of peaks attributed to hydrogen bonds for visualization purposes, showing the conservation in features between the similar poses but not the incorrect docking pose.}
  \label{fig:si-example}
\end{figure}

\newpage

\begin{figure}[h]
  \centering
  \includegraphics[width=\textwidth,height=1.0\textheight,keepaspectratio]{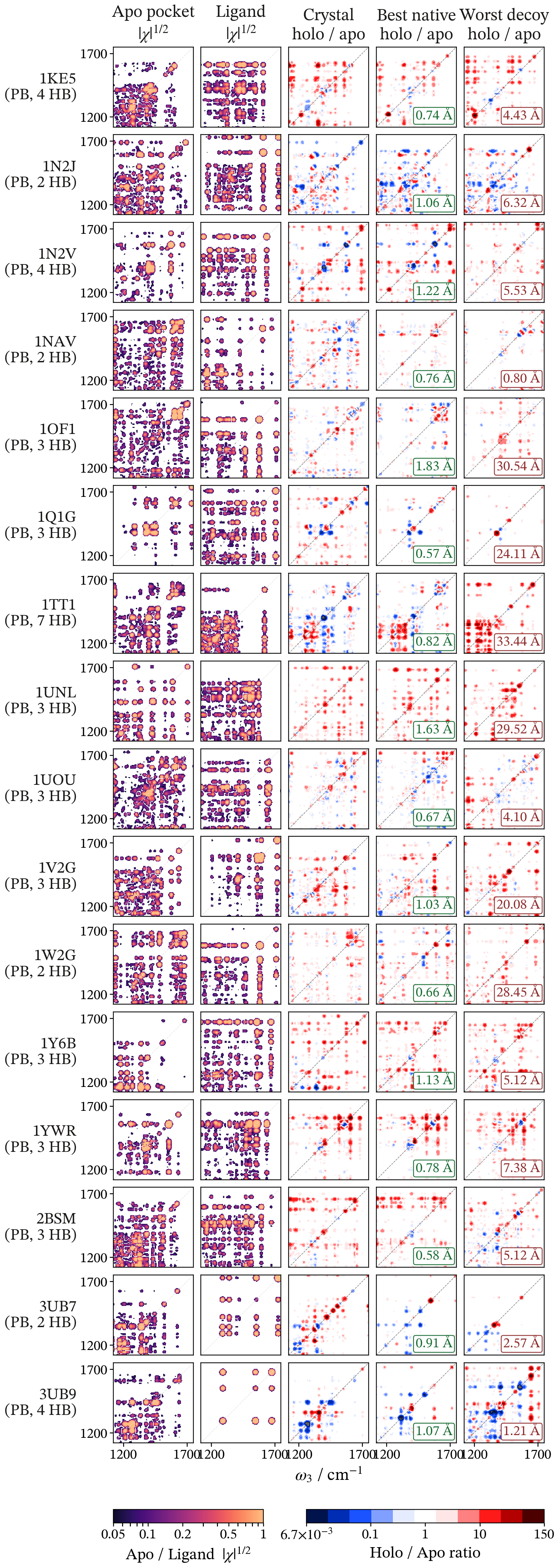}
  \caption{PDB IDs from top to bottom: 1KE5, 1N2J, 1N2V, 1NAV, 1OF1, 1Q1G, 1TT1, 1UNL, 1UOU, 1V2G, 1W2G, 1Y6B, 1YWR, 2BSM, 3UB7, 3UB9}
  \label{fig:si-example}
\end{figure}

\begin{figure}[h]
  \centering
  \includegraphics[width=\textwidth,height=1.0\textheight,keepaspectratio]{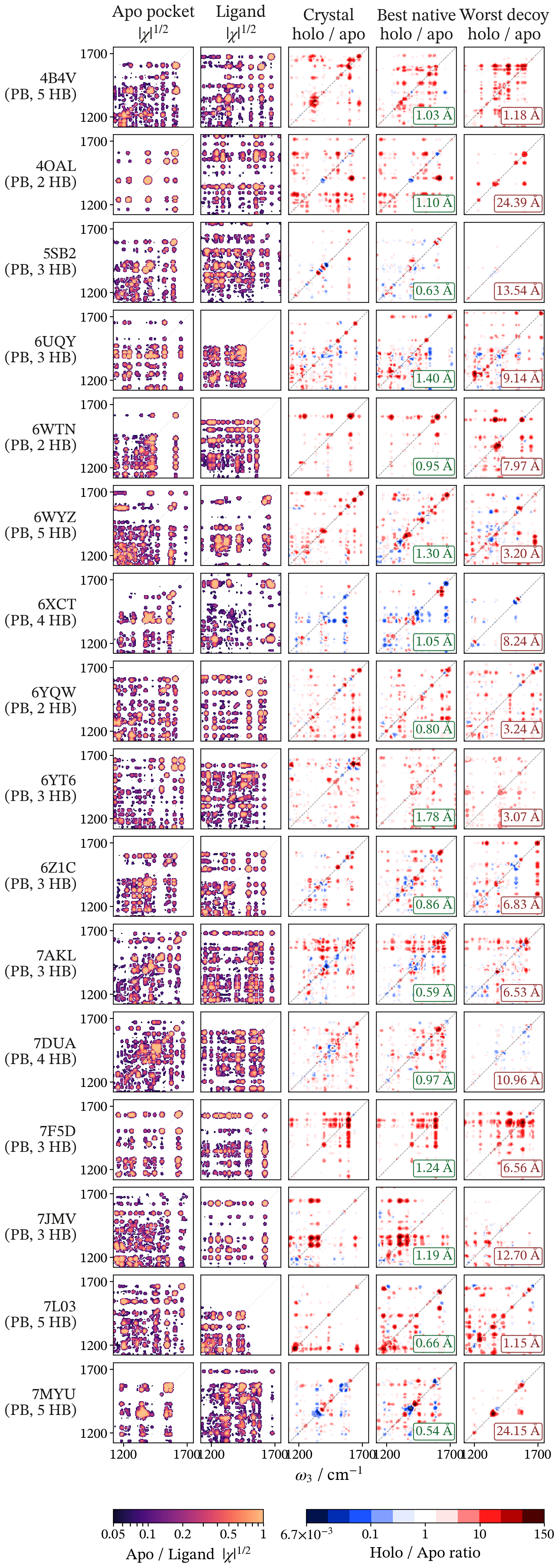}
  \caption{PDB IDs from top to bottom: 4B4V, 4OAL, 5SB2, 6UQY, 6WTN, 6WYZ, 6XCT, 6YQW, 6YT6, 6Z1C, 7AKL, 7DUA, 7F5D, 7JMV, 7L03, 7MYU}
  \label{fig:si-example}
\end{figure}

\begin{figure}[h]
  \centering
  \includegraphics[width=\textwidth,height=1.0\textheight,keepaspectratio]{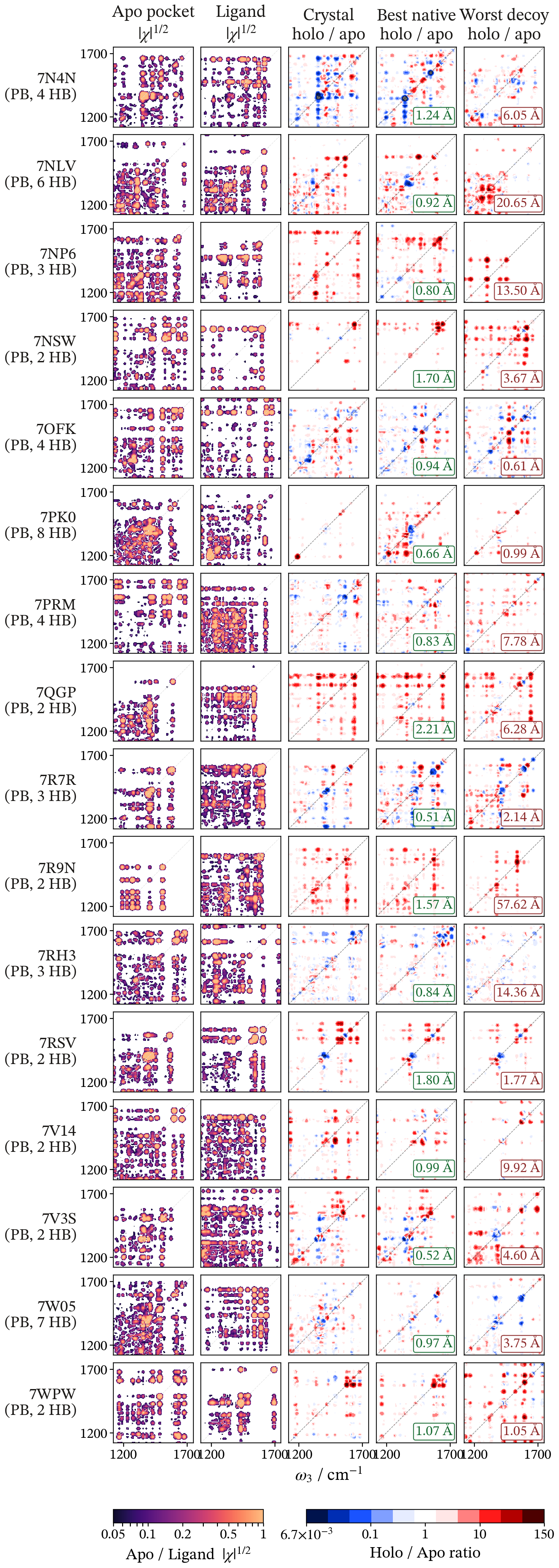}
  \caption{PDB IDs from top to bottom: 7N4N, 7NLV, 7NP6, 7NSW, 7OFK, 7PK0, 7PRM, 7QGP, 7R7R, 7R9N, 7RH3, 7RSV, 7V14, 7V3S, 7W05, 7WPW}
  \label{fig:si-example}
\end{figure}

\begin{figure}[h]
  \centering
  \includegraphics[width=\textwidth,height=1.0\textheight,keepaspectratio]{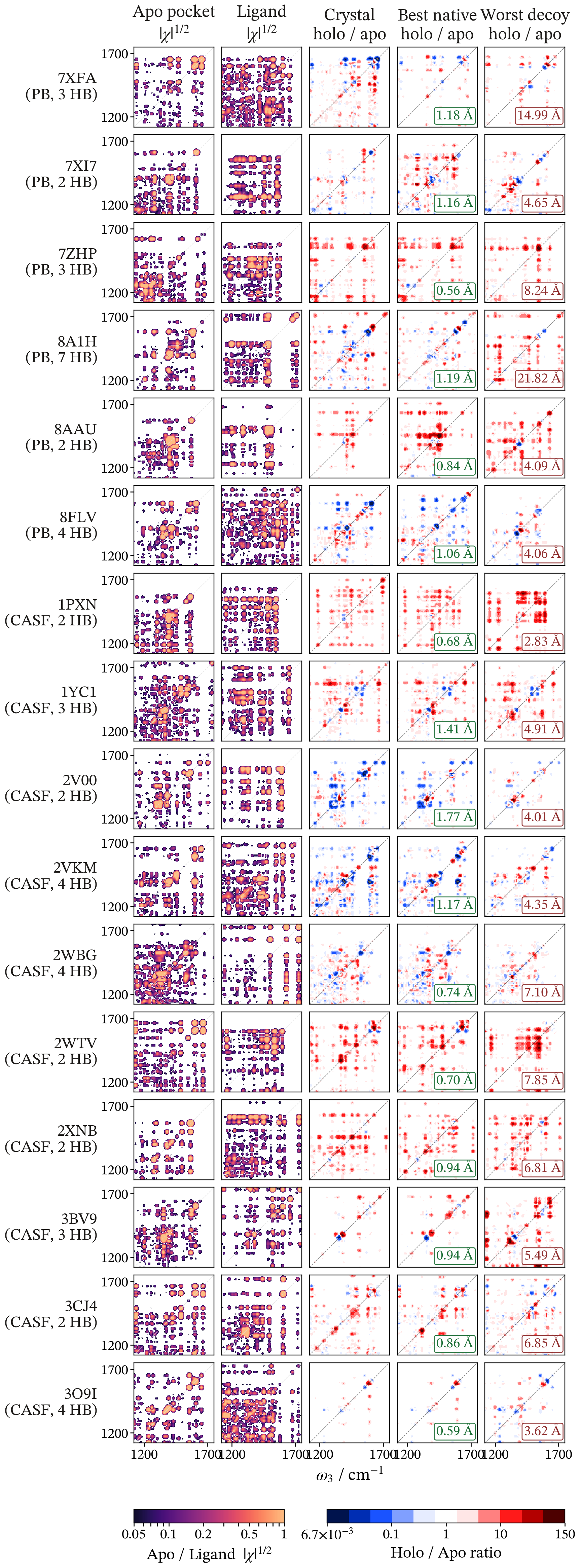}
  \caption{PDB IDs from top to bottom: 7XFA, 7XI7, 7ZHP, 8A1H, 8AAU, 8FLV, 1PXN, 1YC1, 2V00, 2VKM, 2WBG, 2WTV, 2XNB, 3BV9, 3CJ4, 3O9I}
  \label{fig:si-example}
\end{figure}

\begin{figure}[h]
  \centering
  \includegraphics[width=\textwidth,height=1.0\textheight,keepaspectratio]{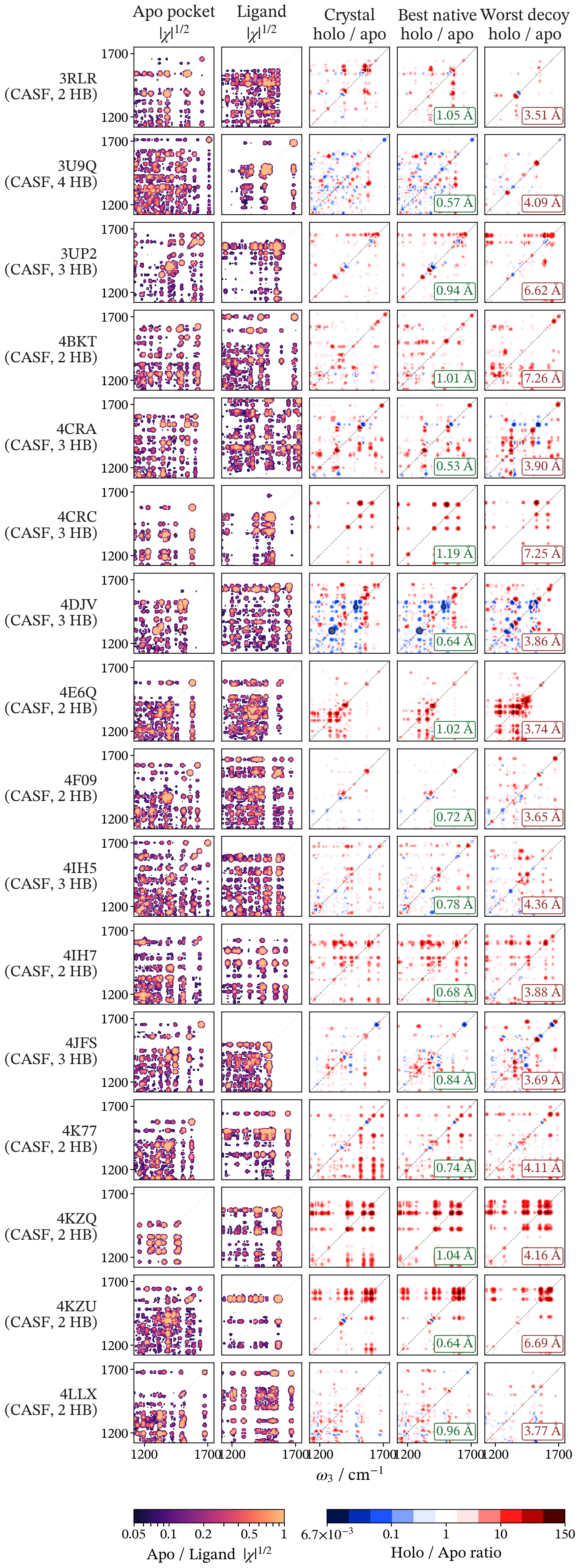}
  \caption{PDB IDs from top to bottom: 3RLR, 3U9Q, 3UP2, 4BKT, 4CRA, 4CRC, 4DJV, 4E6Q, 4F09, 4IH5, 4IH7, 4JFS, 4K77, 4KZQ, 4KZU, 4LLX}
  \label{fig:si-example}
\end{figure}

\begin{figure}[h]
  \centering
  \includegraphics[width=\textwidth,height=1.0\textheight,keepaspectratio]{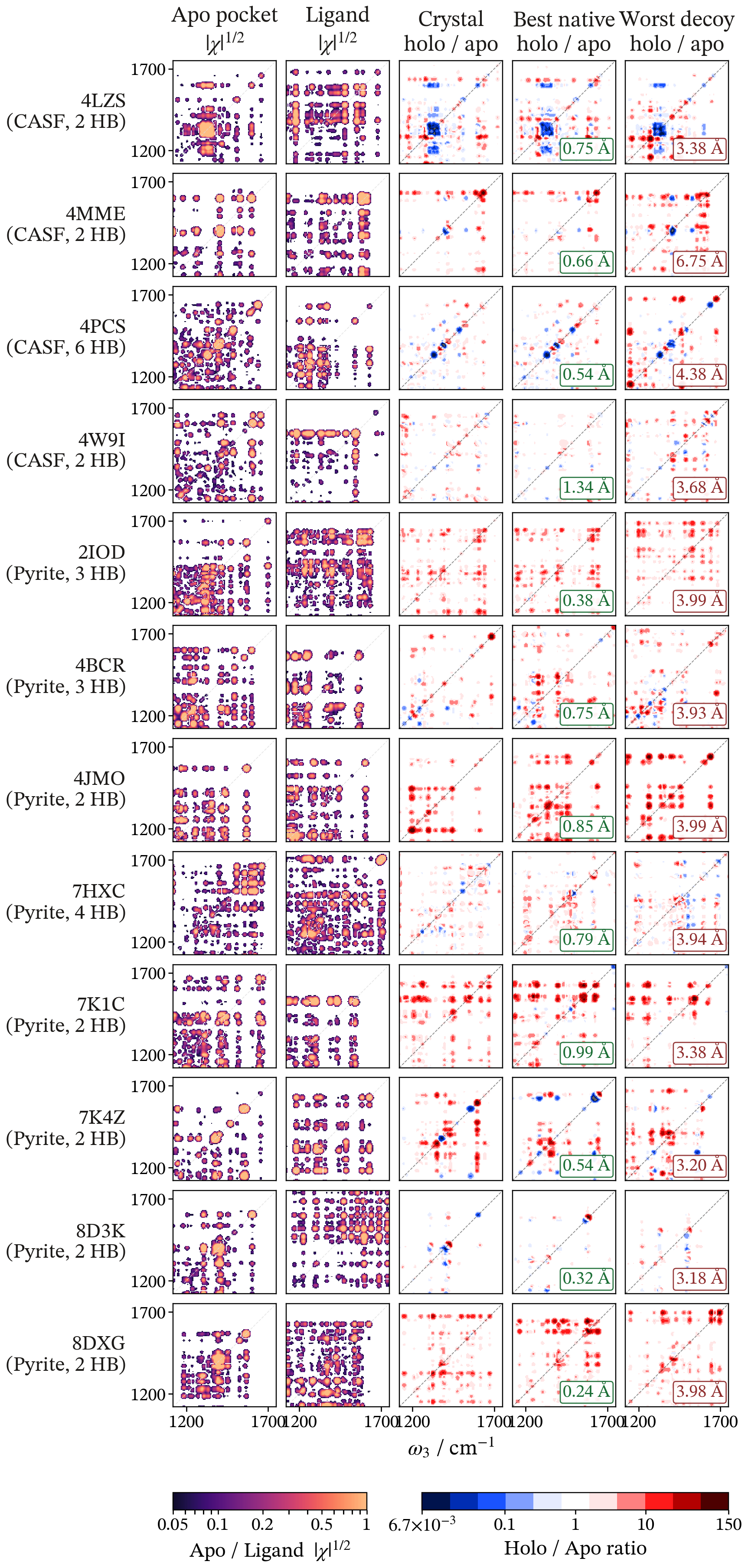}
  \caption{PDB IDs from top to bottom: 4LZS, 4MME, 4PCS, 4W9I, 2IOD, 4BCR, 4JMO, 7HXC, 7K1C, 7K4Z, 8D3K, 8DXG}
  \label{fig:si-example}
\end{figure}

\end{document}